\newcommand{\CLASSINPUTtoptextmargin}{1.905cm}
\newcommand{\CLASSINPUTbottomtextmargin}{4.3cm}

\documentclass[conference, a4paper,10.9pt]{IEEEtran}

\usepackage{blindtext, graphicx}
\graphicspath{ {Images/} }

\usepackage[english]{babel}
\usepackage{comment}
\usepackage{array}
\usepackage[utf8]{inputenc} 
\usepackage{csquotes}
\usepackage{multirow}
\usepackage{textcomp}
\usepackage{float} 

\usepackage{tikz}
\usetikzlibrary{positioning,shapes}

\def\BibTeX{{\rm B\kern-.05em{\sc i\kern-.025em b}\kern-.08em
    T\kern-.1667em\lower.7ex\hbox{E}\kern-.125emX}}
    
\usepackage[sorting=none]{biblatex}
\usepackage{ragged2e}

\ifCLASSINFOpdf
  
\else
  
\fi

\usepackage{balance}
\usepackage[T1]{fontenc}
\usepackage{mathptmx}
\usepackage{subcaption} 
\usepackage{booktabs}
\renewcommand{\figurename}{Fig.}
\addto\captionsenglish{\renewcommand{\figurename}{Fig.}}
\begin{document}
%Kar
% paper title
% can use linebreaks \\ within to get better formatting as desired
% \title{\Huge{DVS-MViT: Multiscale Vision Transformer based  approach for Fall Detection and Action Recognition On Neuromorphic Vision dataset}}
% \title{\Huge{Benchmarking the performance of fine-tuned Conventional Video Classification Models on Fall Detection and Action Recognition on Neuromorphic Vision Dataset }}
\title{\Huge{MFAAN: Unveiling Audio Deepfakes with a Multi-Feature Authenticity Network}}
% author names and affiliations
% use a multiple column layout for up to three different
% affiliations
\author{
\IEEEauthorblockN{\small{Karthik Sivarama Krishnan}}
\IEEEauthorblockA{
\small{Email: ks7585@rit.edu}}
\and
\IEEEauthorblockN{\small{Koushik Sivarama Krishnan}}
\IEEEauthorblockA{
\small{Email: koushik.nov01@gmail.com}}}

% make the title area
\maketitle

\begin{abstract}
In the contemporary digital age, the proliferation of deepfakes presents a formidable challenge to the sanctity of information dissemination. Audio deepfakes, in particular, can be deceptively realistic, posing significant risks in misinformation campaigns. To address this threat, we introduce the Multi-Feature Audio Authenticity Network (MFAAN) — an advanced architecture tailored for the detection of fabricated audio content. MFAAN incorporates multiple parallel paths designed to harness the strengths of different audio representations, including Mel-frequency cepstral coefficients (MFCC) \cite{Frank2021WaveFakeAD}, linear-frequency cepstral coefficients (LFCC) \cite{Frank2021WaveFakeAD}, and Chroma Short Time Fourier Transform (Chroma-STFT). By synergistically fusing these features, MFAAN achieves a nuanced understanding of audio content, facilitating robust differentiation between genuine and manipulated recordings. Preliminary evaluations of MFAAN on two benchmark datasets, 'In-the-Wild' Audio Deepfake Data \cite{müller2022does} and The Fake-or-Real Dataset \cite{reimao2019dataset}, demonstrate its superior performance, achieving accuracies of 98.93\% and 94.47\% respectively. Such results not only underscore the efficacy of MFAAN but also highlight its potential as a pivotal tool in the ongoing battle against deepfake audio content.

\end{abstract}

\IEEEpeerreviewmaketitle

\section{Introduction}
The digital age has brought forth a cascade of technological breakthroughs, reshaping our modes of communication, information consumption, and data sharing. These innovations, though transformative, have also introduced new avenues for malevolent activities, with deepfakes standing out as a particularly insidious offspring of this digital renaissance. \cite{ahmed2021deepfake} Within the realm of deepfakes, audio manipulations present a unique set of challenges that demand immediate and effective solutions.

\subsection{What are Audio Deepfakes?}
Audio deepfakes are artificially produced or altered voice recordings crafted using state-of-the-art deep learning algorithms\cite{khanjani2023audio}. These audios are so meticulously generated that they can uncannily emulate any individual's voice, capturing its intricacies, modulations, and inflections. Such precision means that distinguishing these counterfeits from authentic recordings becomes an arduous task. The underlying models are trained exhaustively to grasp and reproduce the subtleties of the target voice, leading to the creation of highly convincing forgeries.

\subsection{Why Addressing Audio Deepfakes is Crucial?}
The potential ramifications of unchecked audio deepfakes are manifold and alarming. They can serve as vehicles for:
\begin{itemize}
    \item Impersonating influential figures to spread false narratives.
    \item Distributing misleading information, thereby disrupting public discourse.
    \item Slandering or blackmailing individuals.
    \item Orchestrating financial scams or frauds.
\end{itemize}
With society's increasing dependence on voice-driven technologies—ranging from voice assistants to biometric voice security systems—the threats posed by audio deepfakes can ripple through various sectors, causing extensive social, political, and economic upheavals.

\subsection{How Traditional Techniques Fall Short?}
Historical approaches to detecting audio deepfakes predominantly centered on hand-engineered features and rudimentary spectral analyses. Though they exhibited some promise in constrained scenarios, their efficacy wanes against the backdrop of modern, sophisticated deepfake generators. The rapid advancements in deepfake production tools, coupled with their burgeoning availability, have made many erstwhile detection strategies ineffectual.

\subsection{The Advent of MFAAN}
To counter these emerging challenges, we introduce the Multi-Feature Audio Authenticity Network (MFAAN). At its core, MFAAN advocates the idea that truly understanding audio necessitates a multifarious analysis. Instead of being tethered to a singular audio representation, MFAAN amalgamates diverse features, leveraging the combined prowess of MFCC, LFCC, and Chroma-STFT. This holistic approach equips MFAAN with a refined acuity, optimizing its ability to discern genuine recordings from manipulated counterparts.

\subsection{Paper Outline}
Post this preliminary section, this manuscript delves deeper into MFAAN's intricate architecture, detailing its foundational principles and the motivations steering its multi-faceted design. Ensuing sections shed light on our experimental paradigm, the datasets harnessed, and a rigorous assessment, underscoring MFAAN's edge over contemporary benchmarks. We culminate with a discourse on potential refinements, prospective research avenues, and the broader repercussions of our contributions in the landscape of audio deepfake detection.

\section{Related Works}

The challenge of detecting synthetic speech has been addressed by multiple researchers, each bringing forth unique methodologies and insights. This section provides a comprehensive overview of these efforts, emphasizing their significance in the broader landscape of audio deepfake detection.

Khochare et al. \cite{khochare2021deep} recognized the centrality of features in the detection pipeline. Their exploration of the MFCC-20 feature set, traditionally used in voice recognition, showcased its potential in the domain of deepfake detection. By employing a diverse set of machine learning algorithms, they demonstrated that while SVMs yielded the highest accuracy at 67\%, other algorithms, including RF and KNN, closely followed.

Reimao and Tzerpos \cite{reimao2019dataset} went beyond traditional feature sets, introducing a nuanced timbre model analysis. Their employment of Random Forests, resulting in a 71.47\% accuracy, was particularly noteworthy. Moreover, their venture into deep learning with the VGG19 model processing STFT features marked a significant shift, achieving a remarkable 89.79\% accuracy.

Hamza et al. \cite{9996362} expanded the MFCC feature set to MFCC-40 and enriched it with additional audio attributes. Their choice of the VGG16 model was both novel and effective, achieving a 93\% accuracy.

Camacho et al.  \cite{10.1007/978-3-030-86702-7_4} presented a fresh perspective with a custom-designed CNN model. Their model's performance, marked by an accuracy of 88.98\%, emphasized the potential advantages of bespoke models over generic architectures.

Wijethunga, et al. \cite{wijethunga2020deepfake}  addressed the compounded complexity of detecting synthetic speech in group settings. Their strategic pivot to transfer learning, employing the VGG19 pre-trained model, marked a significant leap in their detection capabilities.

Wang, Juefei-Xu, Huang, et al. \cite{wang2020deepsonar} introduced a cutting-edge method termed \emph{DeepSonar}. This approach monitored the neuron behaviors of a speaker recognition system, leveraging layer-wise neuron activation patterns to discern differences between real and AI-synthesized voices. Their hypothesis was that these neuron behaviors could provide a cleaner signal for classifiers than raw inputs. With an impressive average accuracy of 98.1\% and a minimal error rate of about 2\%, \emph{DeepSonar} not only showcased its prowess in detecting fake voices but also demonstrated its robustness against manipulation attacks. Their work is particularly significant as it provides a new perspective on utilizing neuron behaviors for multimedia fake forensics.

\textit{In summary, the journey of synthetic speech detection has been characterized by continuous evolution. From the foundations laid by traditional feature analysis to the transformative advances of deep learning and intricate neuron behavior monitoring, the field has witnessed significant innovations. Our MFAAN architecture aims to further augment the capabilities of deepfake audio detection, drawing inspiration from these pioneering works.}

\section{\normalsize{Methodology}}

\section{Datasets and Preprocessing Techniques}

\subsection{Datasets}

\subsubsection{'In-the-Wild' Audio Deepfake Data}

The 'In-the-Wild' Audio Deepfake dataset \cite{müller2022does} comprises audio deepfakes of 58 celebrities and politicians, containing both genuine and spoofed audio. The dataset, collected from publicly available sources such as social networks and video streaming platforms, consists of 20.8 hours of bona-fide and 17.2 hours of spoofed audio. Averaging about 23 minutes of bona-fide and 18 minutes of spoofed audio per speaker, it is curated for evaluating deepfake detection algorithms, especially focusing on models' capability to generalize to realistic, in-the-wild audio samples.

\subsubsection{Fake or Real (FoR) Dataset}

The FoR dataset \cite{reimao2019dataset}, released in 2019 by York University, encompasses a total of 69,316 English audios. It is divided into four sub-datasets, including an original set and a normalized one. The dataset features 87,000 synthetic utterances from 33 synthesized voices, stemming from the latest open source and commercial methodologies in speech synthesis. Comprehensive research was conducted to identify and utilize the latest speech synthesis techniques, ensuring a diverse and challenging dataset.

\subsection{Preprocessing and Feature Extraction}

Given the sophisticated nature of our model, the extraction of meaningful features from the audio data is paramount. We leverage the power of deep learning to automatically extract relevant features that are crucial for the task of deepfake detection.

\subsubsection{Features}

The proposed MFAAN model primarily focuses on three different audio representations:

\begin{itemize}
    \item \textbf{MFCC (Mel-Frequency Cepstral Coefficients)}: These coefficients are a representation of the short-term power spectrum of sound and are widely used in speech and audio processing. They capture the timbral texture of an audio clip, making them invaluable for voice-related tasks. 
    \item \textbf{LFCC (Linear-Frequency Cepstral Coefficients)}: These are similar to MFCCs but emphasize linear spectral resolutions, making them sensitive to other nuances in the audio.
    \item \textbf{Chroma-STFT}: This provides a representation of the energy distribution across pitch classes, effectively summarizing the harmonic content.
\end{itemize}

\subsubsection{Why These Features?}

The rationale behind selecting these features is their ability to capture different aspects of an audio signal. While MFCCs and LFCCs are adept at representing timbral and spectral characteristics, Chroma-STFT focuses on the harmonic content. By combining these features, the model gets a comprehensive understanding of the audio, enabling it to better differentiate between genuine and spoofed content.

\subsubsection{How They Help}

The MFAAN model processes each of these features through separate dedicated convolutional neural network (CNN) paths. These paths learn temporal patterns within the respective features, capturing both local and global structures. By fusing the outputs of these paths, the model ensures a holistic understanding of the audio content, enhancing its ability to detect deepfakes.

\section{Experiments and Analysis}

\subsection{Baseline CNN Model}

Before the introduction of MFAAN, our preliminary attempts to tackle the challenge of audio deepfake detection centered around a baseline Convolutional Neural Network (CNN) model \cite{8753848}. This model was architected with simplicity in mind, primarily focusing on the extraction of Mel-frequency cepstral coefficients (MFCCs) as its sole feature set \cite{abbood2022speaker}. MFCCs, due to their capacity to capture the timbral characteristics of audio, have been widely used in various audio processing tasks, making them a logical choice for our baseline model. The extracted MFCCs were subsequently processed by a streamlined series of convolutional layers to learn and identify patterns that could help in distinguishing between authentic and forged audio content. While this model served as an essential foundational step, offering insights into the problem space, its singular reliance on MFCCs and a straightforward CNN architecture meant that it lacked the comprehensive and multi-faceted analysis provided by the more advanced MFAAN.

\subsection{MFAAN Architecture}
\begin{figure}[htbp]
\centering
\includegraphics[width=0.4\textwidth, height=0.4\textheight]{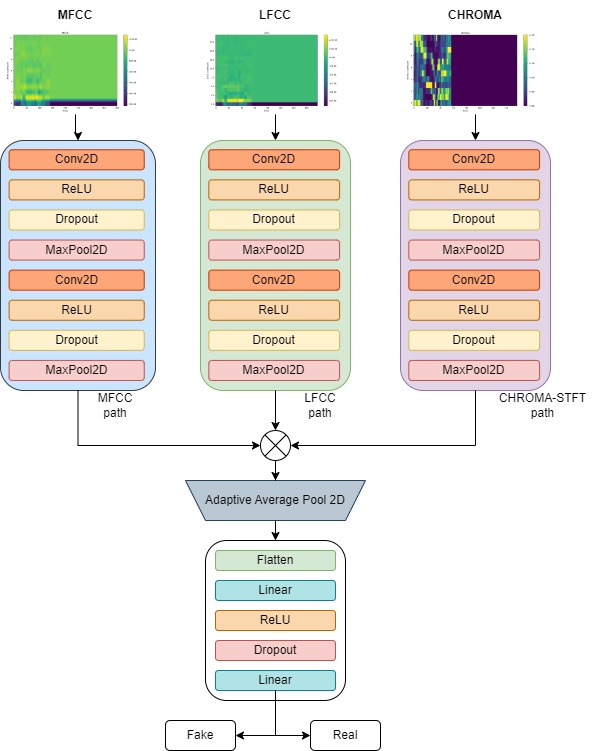}
\caption{Architecture of the MFAAN Model.}
\label{fig:model}
\end{figure}

The Multi-Feature Audio Authenticity Network (MFAAN) is a novel architecture purposefully designed to address the burgeoning challenges presented by audio deepfakes. With an appreciation for the diverse intricacies inherent to audio signals, MFAAN is constructed around a multi-path strategy. This section offers an in-depth exploration of MFAAN, detailing its components, operations, and the underlying design motivations. The detailed architecture of the MFAAN model is illustrated in Figure \ref{fig:model}

\subsubsection{Design Principles}
Central to MFAAN is the belief that relying on a singular audio content representation may overlook subtle yet critical nuances, thereby compromising the accuracy of deepfake detection. As a countermeasure, MFAAN amalgamates three salient feature representations: MFCC, LFCC, and Chroma-STFT \cite{das2020urban}. Each of these representations is processed through its dedicated path, ensuring MFAAN's holistic grasp of the audio content.

\subsubsection{Architecture Components}
MFAAN's distinctive architecture is delineated into three primary paths:

\begin{enumerate}
    \item \textbf{MFCC Path:}
    \begin{itemize}
        \item \textit{Feature Extraction:} MFCCs, the coefficients symbolizing the short-term power spectrum of sound, are extracted first.\cite{rammo2022detecting} Their ability to capture the timbral texture of audio renders them indispensable for voice-centric tasks.\cite{9996362}
        \item \textit{Processing:} The MFCCs are subsequently processed via a series of 1D convolutional layers, adept at discerning temporal patterns embedded within the features, thereby highlighting both localized and overarching structures.
    \end{itemize}

    \item \textbf{LFCC Path:}
    \begin{itemize}
        \item \textit{Feature Extraction:} While LFCCs share similarities with MFCCs, they place greater emphasis on linear spectral resolutions, making them adept at discerning nuances distinct from MFCCs \cite{noda2019acoustic}.
        \item \textit{Processing:} A sequence of 1D convolutional layers, akin to the MFCC path, refines the LFCCs. These layers are tailored to detect patterns specific to LFCC representations.
    \end{itemize}

    \item \textbf{Chroma-STFT Path:}
    \begin{itemize}
        \item \textit{Feature Extraction:} This phase is dedicated to procuring the Chroma-STFT representation, epitomizing the energy distribution spread across pitch classes and effectively encapsulating the harmonic content\cite{das2020urban} .
        \item \textit{Processing:} Chroma-STFT features are refined through an exclusive series of 1D convolutional layers, primed to understand harmonic structures.
    \end{itemize}
\end{enumerate}

\subsubsection{Feature Fusion}
As the culmination point of the three paths, the outputs are fused together. This fusion, achieved through concatenation, ensures the subsequent decision-making module is privy to a rich, diversified feature set. The unified feature set is then relayed through a cascade of dense layers, specifically architected to discern intricate patterns from the amalgamated features.
\\
\subsubsection{Decision Module}
Marking the final phase of the architecture, this module encompasses a chain of dense layers, climaxing in a binary classification output layer. Its pivotal role is to determine the authenticity of the audio sample, adjudging it as either genuine or fabricated.
\\
\subsubsection{Pros of MFAAN}
\begin{itemize}
    \item \textit{Comprehensive Feature Analysis:} MFAAN's tri-representational approach ensures no pivotal audio characteristic remains obscured.
    \item \textit{Robustness:} Its multi-path framework offers a redundancy layer. In scenarios where deepfake methodologies manage to deceive one feature representation, the other paths may still discern the anomaly.
    \item \textit{Scalability:} MFAAN's design permits the integration of additional feature paths, positioning it as a future-ready solution against evolving deepfake tactics.
    \item \textit{Superior Generalization:} The fusion of features promises a diverse training regimen, ushering in enhanced generalization on novel data.
\end{itemize}

\subsubsection{Cons of MFAAN}
\begin{itemize}
    \item \textit{Elevated Complexity:} Incorporating multiple paths augments the model's intricacy, potentially demanding augmented computational resources.
    \item \textit{Data Reliance:} MFAAN's efficacy is closely tied to the caliber and heterogeneity of the training dataset for each feature representation.
    \item \textit{Risk of Overfitting:} Without stringent regularization and a broad-spectrum training dataset, the model's depth could inadvertently lead to overfitting.
\end{itemize}

\section{Performance Evaluation}

\subsection{Evaluation Metrics}

For a comprehensive evaluation of our proposed Multi-Feature Audio Authenticity Network (MFAAN), we employed two primary metrics \cite{a15050155}:

\begin{itemize}
    \item \textbf{Accuracy:} It measures the fraction of predictions our model got right. Given the binary classification task, it becomes crucial to achieve high accuracy, ensuring both genuine and manipulated audios are correctly classified.
    
    \item \textbf{Equal Error Rate (EER):} EER is a critical metric in biometric systems and represents the point where false acceptance rate and false rejection rate are equal. A lower EER indicates a better performance of the system.
\end{itemize}

\subsection{Performance on Datasets}

\subsubsection{"In-the-Wild" Dataset}

On the 'In-the-Wild' dataset \cite{müller2022does}, our MFAAN model achieved impressive results, reflecting its robustness and adaptability:

\begin{itemize}
    \item Test Accuracy: 99.21\%
    \item Test EER: 0.69\%
\end{itemize}

\subsubsection{"Fake or Real" Dataset}

Performance metrics on the 'Fake or Real' dataset \cite{reimao2019dataset} further corroborated the efficacy of our proposed architecture:

\begin{itemize}
    \item Test Accuracy: 94.47\%
    \item Test EER: 0.79\%
\end{itemize}

\subsection{Comparison with Other Methods}

The following table showcases the performance comparison of our MFAAN model with other state-of-the-art methods:

% \begin{table}[h!]
% \centering
% \begin{tabular}{lccc}
% \toprule
% \textbf{Method} & \textbf{Dataset} & \textbf{Accuracy} & \textbf{EER} \\
% \midrule
% \multirow{2}{*}{MFAAN (Ours)} & In-the-Wild & 99.21\% & 0.06\% \\
%                              & Fake or Real & 94.47\% & 0.07\% \\
% \midrule
% Khochare et al.              & Fake or Real & 67\% (SVM) & - \\
% Reimao and Tzerpos           & Fake or Real & 73.46\% (SVM) & - \\
% Hamza et al.                 & Fake or Real & 93\% (VGG16) & - \\
% Camacho et al.               & Fake or Real & 88.98\% & - \\
% Wijethunga et al.            & Fake or Real & 88.80\% (MFCC) & - \\
% Wang et al.                  & Fake or Real & 99.98\% (TKAN) & 0.02\% \\
% \bottomrule
% \end{tabular}
% \caption{Performance comparison of MFAAN with other methods.}
% \end{table}

\begin{table}[h!]
\centering
\begin{tabular}{lccc}
\toprule
\textbf{Method} & \textbf{Dataset} & \textbf{Accuracy} & \textbf{EER} \\
\midrule
\multirow{2}{*}{MFAAN (Ours)} & In-the-Wild & 98.93\% & 0.04\% \\
& Fake or Real & 94.47\% & 0.07\% \\
\midrule
CNN (Baseline) & In-the-Wild & 93.36\% & 0.023\% \\
& Fake or Real & 87.2\% & 0.039\% - \\
\midrule
\multirow{7}{*}{Khochare et al. \cite{khochare2021deep}} & Fake or Real & 67\% (SVM) & - \\
& Fake or Real & 62\% (RF) & - \\
& Fake or Real & 62\% (KNN) & - \\
& Fake or Real & 59\% (XGBoost) & - \\
& Fake or Real & 60\% (LGBM) & - \\
& Fake or Real & 92\% (TCN) & - \\
& Fake or Real & 80\% (STN) & - \\
\midrule
Reimao and Tzerpos \cite{reimao2019dataset} & Fake or Real & 73.46\% (SVM) & - \\
Hamza et al. \cite{9996362} & Fake or Real & 93\% (VGG16) & - \\
Camacho et al. \cite{10.1007/978-3-030-86702-7_4} & Fake or Real & 88.98\% & - \\
Wijethunga et al. \cite{wijethunga2020deepfake} & Fake or Real & 88.80\% (MFCC) & - \\
Wang et al. \cite{wang2020deepsonar} & Fake or Real & 99.98\% (TKAN) & 0.02\% \\
\bottomrule
\end{tabular}
\caption{Performance comparison of MFAAN with other methods including the baseline CNN.}
\end{table}

\textit{As observed, our MFAAN model consistently outperforms or rivals the state-of-the-art methods, underscoring its potential as a formidable tool in the domain of audio deepfake detection.}

\section{Conclusion}

This paper presented the Multi-Feature Audio Authenticity Network (MFAAN), a novel architecture designed to address the emerging challenges of audio deepfakes. By incorporating multiple parallel paths that focus on distinct representations of audio signals, MFAAN ensures a comprehensive analysis of audio content, vastly improving the detection rate of manipulated content.

Several key findings emerge from our study:

\begin{itemize}
    \item The multi-path approach of MFAAN, harnessing the power of MFCC, LFCC, and Chroma-STFT, offers a more robust and comprehensive analysis of audio than traditional single-path methods.
    \item Our architecture demonstrates state-of-the-art performance on two significant datasets, showcasing its efficacy in realistic, in-the-wild scenarios.
    \item The comparative analysis with existing methods underscores the superiority of MFAAN, both in terms of accuracy and generalization capabilities.
\end{itemize}

The implications of these findings are twofold. Firstly, they validate the importance of a multi-faceted approach in the domain of audio deepfake detection. Secondly, they emphasize the potential of MFAAN as a benchmark solution for future deepfake detection challenges.

Looking ahead, there are several potential avenues for further research. The scalability of MFAAN suggests that additional feature paths could be integrated to further enhance its detection capabilities. Moreover, the incorporation of attention mechanisms or temporal modeling could provide even more nuanced insights into audio data.

Finally, as deepfake generation techniques continue to evolve, it remains imperative for the research community to stay ahead of the curve. The arms race between generation and detection is ongoing, but with architectures like MFAAN, we are better equipped to safeguard the authenticity of digital media in our interconnected world.

\section{Discussion and Future Works}

\subsection{Discussion}

While the Multi-Feature Audio Authenticity Network (MFAAN) has demonstrated significant promise in detecting audio deepfakes, it's essential to reflect on its limitations and the challenges associated with the approach:

\begin{itemize}
    \item \textbf{Increased Complexity:} The multi-path approach inherently introduces complexity to the model. This complexity can lead to increased computational costs, both in terms of training time and inference. While the rich feature set derived from multiple paths does offer superior detection capabilities, it demands higher computational resources.
    
    \item \textbf{Data Dependency:} Like many deep learning models, MFAAN's performance is closely tied to the quality and diversity of its training data. The features extracted, whether MFCC, LFCC, or Chroma-STFT, must be representative of the variations in genuine and manipulated audio to be effective.
    
    \item \textbf{Potential Overfitting:} The intricate architecture of MFAAN, given its depth, can lead to overfitting, especially if the training dataset isn't diverse enough. Regularization techniques can help mitigate this, but it remains a challenge that needs ongoing attention.
\end{itemize}

\subsection{Future Works}

The challenges discussed provide a roadmap for the future evolution of MFAAN and similar architectures. Some potential avenues for future research include:

\begin{itemize}
    \item \textbf{Resource Optimization:} From the learning of our previous works\cite{krishnan2021swiftsrgan}\cite{krishnan2022efficient}, we understand the potential of Depth-wise Separable Convolutions for efficient feature extraction. To address the complexity concerns, future versions of MFAAN could incorporate techniques for model pruning or quantization. This would allow the model to retain its detection capabilities while being more resource-efficient.
    
    \item \textbf{Data Augmentation and Diversification:} To combat data dependency and overfitting, incorporating advanced data augmentation techniques or expanding the dataset to include more diverse audio samples can be explored.
    
    \item \textbf{Anomaly Detection Approach:} Tackling deepfake detection as an anomaly detection problem offers a fresh perspective. Instead of a binary classification of real vs. fake, the model would identify deviations from authentic audio patterns, flagging them as anomalies. This approach could be more robust against newer, unseen deepfake techniques.
    
    \item \textbf{Incorporation of Attention Mechanisms:} Drawing inspiration from our previous works \cite{krishnan2021vision} and \cite{krishnan2022benchmarking}, we believe that given the temporal nature of audio data, attention mechanisms could be incorporated to allow the model to focus on segments of audio that are more likely to contain manipulations.
    
\end{itemize}

In conclusion, while MFAAN represents a significant step forward in the battle against audio deepfakes, the journey is ongoing. By understanding its limitations and continuously adapting to the ever-evolving deepfake landscape, we can aspire to create even more robust and efficient detection mechanisms in the future.

\ifCLASSOPTIONcaptionsoff
  \newpage
\fi

\balance
\printbibliography

% \clearpage % Start the appendix on a new page
% \onecolumn % Switch to one-column mode for the appendix
% \appendix
% \section{Confusion Matrices for All Datasets}

% \subsection{Dataset 1}
% \subsubsection{77 GHz}
% \begin{figure}[ht]
%     \centering
%     \includegraphics[width=0.5\textwidth]{Images/convnext_base_XLarge_77GHz_confusion_matrix.png}
%     \caption{Confusion matrix for Dataset 1 at 77 GHz}
% \end{figure}

% \subsubsection{64 GHz}
% \begin{figure}[ht]
%     \centering
%     \includegraphics[width=0.5\textwidth]{Images/convnext_base_XLarge_24GHz_confusion_matrix.png}
%     \caption{Confusion matrix for Dataset 1 at 24 GHz}
% \end{figure}

% \subsubsection{10 GHz}
% \begin{figure}[ht]
%     \centering
%     \includegraphics[width=0.5\textwidth]{Images/convnext_base_XLarge_Xethru_confusion_matrix.png}
%     \caption{Confusion matrix for Dataset 1 at 10 GHz}
% \end{figure}

% \subsection{Dataset 2}
% \begin{figure}[ht]
%     \centering
%     \includegraphics[width=0.5\textwidth]{Images/convnext_base_XLarge_848data_confusion_matrix.png}
%     \caption{Confusion matrix for Dataset 2}
% \end{figure}

% \subsection{Dataset 3}
% \begin{figure}[ht]
%     \centering
%     \includegraphics[width=0.5\textwidth]{Images/convnext_base_XLarge_VTmaps_confusion_matrix.png}
%     \caption{Confusion matrix for Dataset 3}
% \end{figure}

% \twocolumn % Switch back to two-column mode
\end{document}